\documentclass[sigconf]{acmart}
\usepackage{multirow}

\AtBeginDocument{%
  \providecommand\BibTeX{{%
    \normalfont B\kern-0.5em{\scshape i\kern-0.25em b}\kern-0.8em\TeX}}}

\setcopyright{acmlicensed}
\begin{document}

\title{Domain-adapted Learning and Imitation: \\
DRL for Power Arbitrage}


\author{Yuanrong Wang}
\affiliation{%
  \institution{University College London}
  \institution{Shell Global Solutions International}
  \country{UK}
}

\author{Vignesh Raja Swaminathan}
\affiliation{%
  \institution{Shell Shared Service Center Chennai}
  \country{India}
}

\author{Nikita P Granger}
\affiliation{%
  \institution{Shell Energy North America}
  \country{USA}
}

\author{Carlos Ros Perez}
\affiliation{%
  \institution{Shell Global Solutions International}
  \country{UK}
}

\author{Christian Michler}
\authornote{Corresponding author.}
\affiliation{%
  \institution{Shell Global Solutions International}
  \country{The Netherlands}
}

\renewcommand{\shortauthors}{Y. Wang et al.}

\begin{abstract}
In this paper, we discuss the Dutch power market, which is comprised of a day-ahead market and an intraday balancing market that operates like an auction. Due to fluctuations in power supply and demand, there is often an imbalance that leads to different prices in the two markets, providing an opportunity for arbitrage. To address this issue, we restructure the problem and propose a collaborative dual-agent reinforcement learning approach for this bi-level simulation and optimization of European power arbitrage trading. We also introduce two new implementations designed to incorporate domain-specific knowledge by imitating the trading behaviours of power traders. By utilizing reward engineering to imitate domain expertise, we are able to reform the reward system for the RL agent, which improves convergence during training and enhances overall performance. Additionally, the tranching of orders increases bidding success rates and significantly boosts profit and loss (P\&L). Our study demonstrates that by leveraging domain expertise in a general learning problem, the performance can be improved substantially, and the final integrated approach leads to a three-fold improvement in cumulative P\&L compared to the original agent. Furthermore, our methodology outperforms the highest benchmark policy by around 50\% while maintaining efficient computational performance.
\end{abstract}




\maketitle

\section{Introduction} \label{intro}

Two main characteristics render the power markets different from the general financial spot market. Firstly, European power markets are mostly energy-only markets where generators are remunerated for generating electric energy instead of capacity, and power storage, e.g. in the form of batteries, is limited and costly. Therefore, the majority of the power markets are not exchange-based real-time spot markets. Instead, forward markets such as day-ahead and intra-day market play an important role. Secondly, electricity markets must continuously balance their supply and demand all the time. The national transmission system operator (TSO) is ultimately responsible for maintaining instantaneous generation-consumption balance. Greater penetration of renewable energy and variation in consumption complicates the process of continuous balancing, leading to moments of surplus/shortage. Arbitrage trading between these markets helps to settle such imbalances.

Arbitrage between two dependent levels of market has been addressed in the literature mostly by bi-level optimization over the last two decades  \citep{Hobbs2000StrategicGA,Bakirtzis2007ElectricityPO,Ruiz2009PoolSO,Kazempour2015StrategicBF,Zou2016PoolEI,Moiseeva2015ExerciseOM,Nasrolahpour2018ImpactsOR,Ye2018InvestigatingTA}.  It addresses an optimization problem under an embedded dependency between two tasks, where one tasks depends on the optimization outcome of the other. Bi-level optimization problems are usually solved after converting them to single-level mathematical programs with equilibrium constraints (MPEC) with equivalent Karush-Kuhn-Tucker (KKT) optimality conditions or linear programming problems. Nevertheless, this modelling framework relies on a non-practical fundamental assumption, i.e. that the underlying problems are continuous and convex \citep{Boyd2006ConvexO}. This limitation is particularly important when modelling markets with complex bidding mechanisms, whose clearing algorithm involves the solution of a multi-period, mixed-integer unit commitment problem, such as many markets in the USA (e.g. CAISO, PJM, NYISO) and the intra-day balancing market in Europe. 

Driven by the advances in artificial intelligence, reinforcement learning (RL) has recently attracted increasing research interest in the power systems community and has emerged as a promising alternative to MPEC formulations in electricity market modeling. Under this framework, RL agents learn to optimize strategies by utilizing experiences acquired from their repeated interactions with the environment, the market clearing process. As learning from experience avoids the derivation of the equivalent KKT optimality conditions, it is capable of addressing the aforementioned challenge of incorporating non-convex operating characteristics into the market bidding process. Previous work employing RL in electricity market modeling \citep{Xiong2002AnES,NaghibiSistani2006ApplicationOQ,Song2000OptimalES,Nanduri2007ARL,Tellidou2007AgentBasedAO,Rahimiyan2010AnA,Yu2010EvaluationOM} has employed conventional Q learning algorithms and its variants \citep{Sutton2005ReinforcementLA}, which rely on look-up tables to approximate the action-value function for each possible state-action pair and thus require discretization of both state and action spaces. Furthermore, a fitted Q-iteration algorithm with kernel-based approximation of the action-value function is proposed in Ormoneit \citep{Ormoneit2017KernelBasedRL}. However, the heuristic choice of the kernel function significantly affects its performance, as it may easily overfit to the history and ignore non-stationarity in the time-series data \citep{Wang2021DynamicPO}. More recently, interest has been growing in deep reinforcement learning (Deep RL), which combines RL with deep learning principles and is driven by the universal function approximation properties of deep neural networks (DNN) \citep{Liang2017WhyDN,Wang2022SparsificationAF}. As an extension of Q-learning using a neural network to map the state space to the action space, Mnih \citep{Mnih2013PlayingAW,Mnih2015HumanlevelCT} proposed the Deep Q Network (DQN) method which employs a DNN to approximate the action-value function and has performed at the level of expert humans in playing Atari 2600 games. Inspired by this pioneering work, several recent papers have employed various Deep RL methods to many applications such as voltage control \citep{Diao2019AutonomousVC}, residential load control \citep{Claessens2018ConvolutionalNN}, building energy management systems \citep{Mocanu2019OnLineBE}, electric vehicles \citep{Wan2019ModelFreeRE}, energy storage scheduling \citep{Boukas2021ADR}, energy trading for prosumers \citep{Chen2018LocalET}, natural gas future spot trading \citep{DeepRLGasReport2022} and process optimization \citep{DeepRLProcessOptReport2022}. 

In this project, arbitrage trading is carried out between the day-ahead market and the balancing market. The participants in the day-ahead market submit bids and offers in advance for the following day. The auction-like balancing market opens before each 15-minute slot to adjust and settle potential outstanding imbalances. The details of the markets will be discussed in Section~\ref{Problems}. As the optimization in the balancing market is also dependant on the position and price from the day-ahead market, this is a bi-level problem, which is modelled by collaborative dual-agent reinforcement learning. Multi-agent RL is challenging due to slow convergence and high dimensionality, see e.g.  \citep{Buoniu2010MultiagentRL,Zhang2019MultiAgentRL}. We designed two novel practical implementations for training and performance enhancement, which are reward engineering based on prior domain knowledge and portfolio tranching for optimized bidding execution, as will be discussed in Section~\ref{IL} and Section~\ref{PT}.

In the remainder of this paper, we present our solutions of a dual-agent deep reinforcement learning framework to solve the bi-level deregulated electricity markets. The details of market composition and structure are presented in Section~\ref{Problems}, and based on that, the derived simulation environment and associated data are discussed in Section~\ref{Methods}. Experimental setup are presented in Section~\ref{Exp} and results are showcased and discussed in Section~\ref{results}. Concluding remarks are presented in Section~\ref{sec:Conclusion}.

\section{Problem Statement}\label{Problems}

\subsection{Power Markets}
 There are various futures markets, addressing time periods from decades in advance, to 15 minutes prior to delivery. Participants can follow load produced by a renewable park and capitalize on price/load variability within the market

Mid/long term contracts are conducted as “over the counter” (OTC) transactions. OTC transactions are responsible for generators and consumers agree on trade contract bilaterally or through a broker, from years up to days before delivery. Regulated exchanges allow parties to submit bids and offers for standardized short-term markets up to 15 minutes before delivery. All positions must balance and clear after day-ahead and intra-day markets, otherwise the imbalance will be settled in the imbalance market. Ancillary services provided to the Transmission System Operator (TSO) are balancing options of last resort.

\begin{figure}[h]
	\centering
	\includegraphics[scale=0.3]{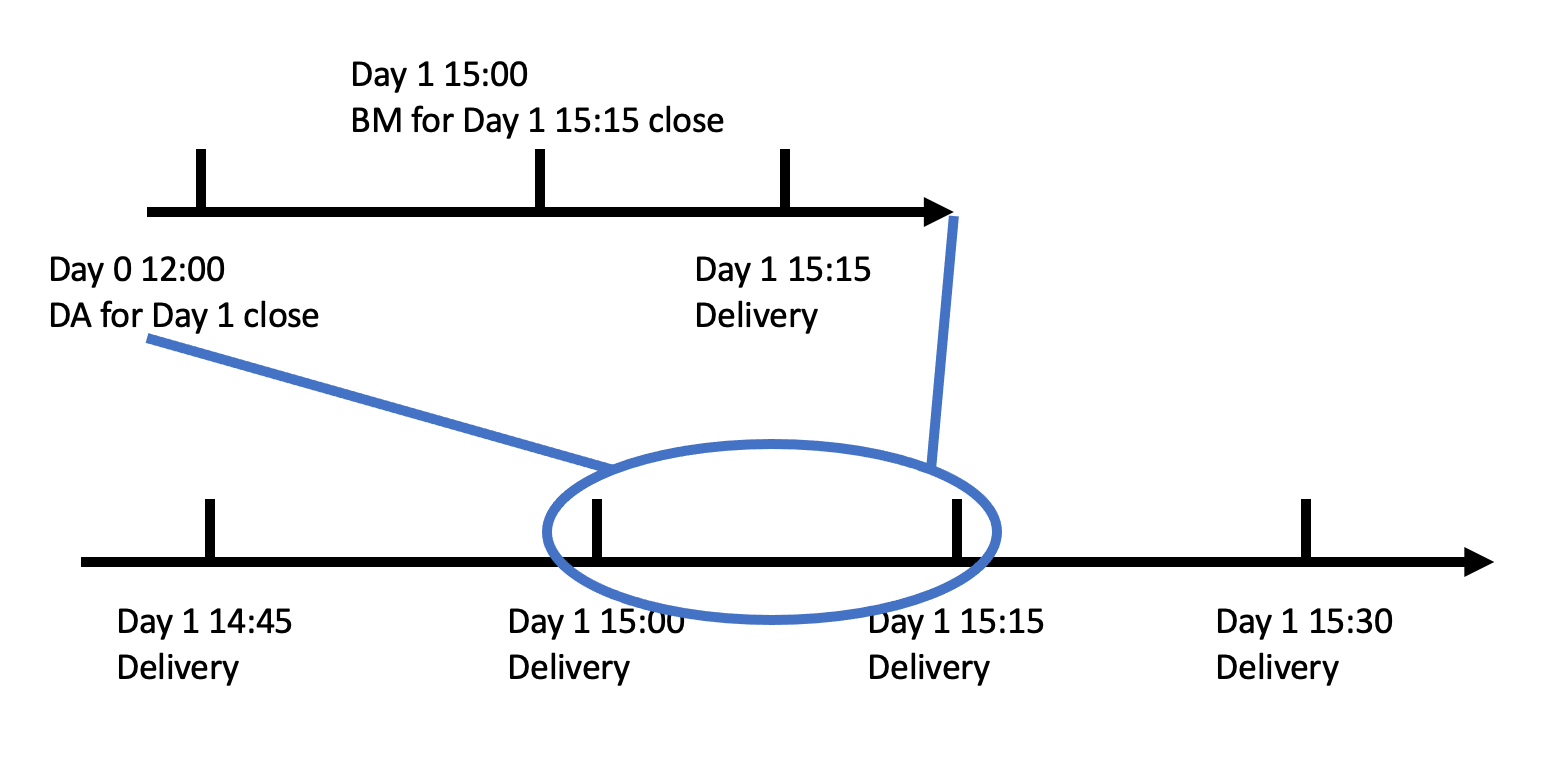}
	\vspace{0.5pt}
	\caption[Market]{Trading and delivery time of day ahead (DA) and balancing market (BM). Day-ahead auction for all hourly slots closes at 12:00 noon the previous day, balancing market closes 15 minutes prior to the delivery time.}\label{fig:market}
\end{figure}

For short-term markets, electricity is traded one day before delivery in the day-ahead market, where all imbalance from the long and mid-term markets must be cleared. However, shifts in day-ahead nominations might still occur due to, e.g., forecast errors. Participants can correct such imbalances in the intra-day market and in the 96 slots of the balancing market (each slot being 15 mins), a detailed schematic is showed in fig~\ref{fig:market}. Any surplus or shortage in the balancing market creates price volatility, and the market rewards parties keeping the grid in equilibrium, e.g., electrolyser day-ahead positions can be sold in the balancing market during shortage and conversely incentivised to consume during surplus. 

\subsection{Bidding in the Balancing Market}
\begin{figure}[h]
	\centering
	\includegraphics[scale=0.375]{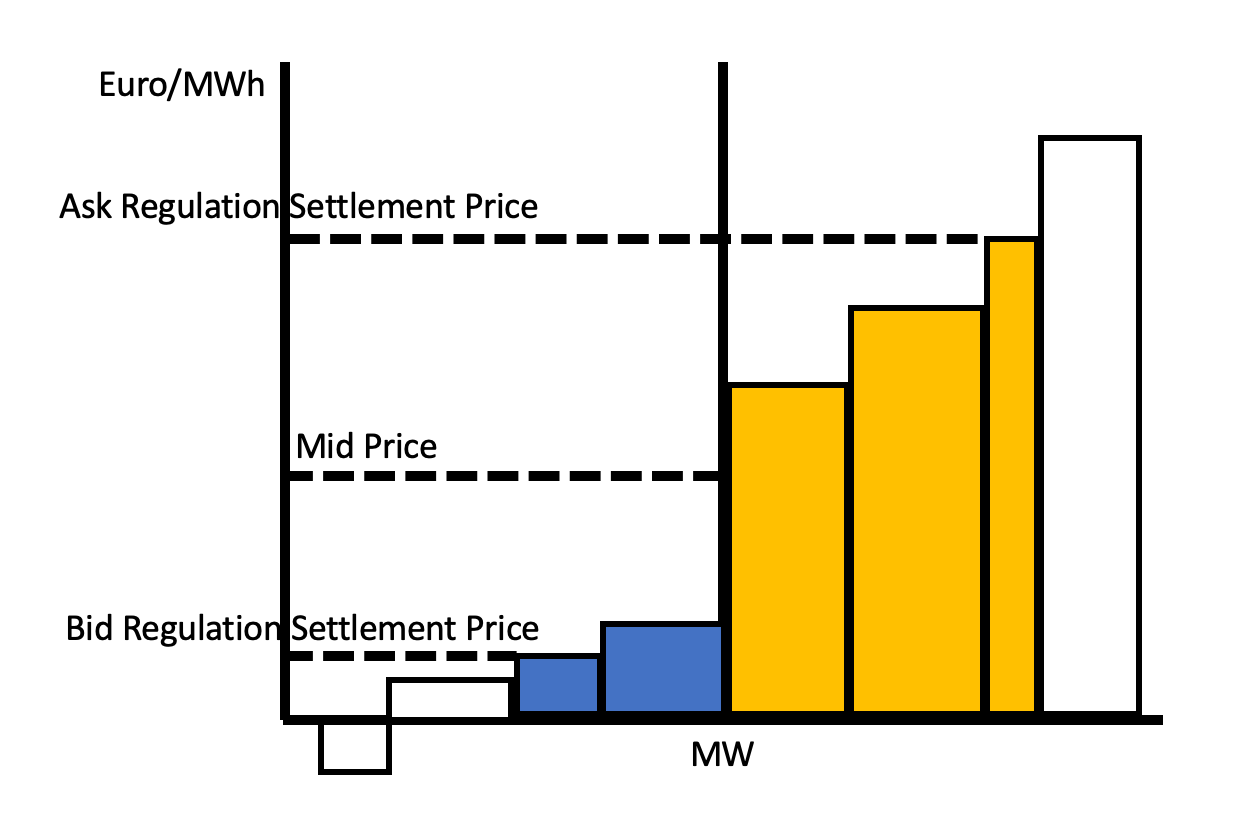}
	\vspace{0.5pt}
	\caption[ladder]{Balancing Market (BM) auction ladder. Only the merit order (blue and yellow colour) within the settlement price range will be accepted by TenneT and settled at the settlement price, other bidding (while colour) will be discarded.}\label{fig:bmladder}
\end{figure}

The balancing market operates as an order-book structured auction, where bids and asks are settled/cleared in a centralized way at the end of the 15-min interval. Hence, bids and asks outside of the current spread (settlement price range) in the 15-minute interval will not be considered, and these bids are discarded, demonstrated by fig~\ref{fig:bmladder}. The settlement price range is determined by the supply and demand of electricity under the sole discretion of TenneT. To increase the success rate of bidding, tranching is used to break a large order into smaller orders across the bid ladder to take advantage of the paid-as-bid nature of capacity auctions.

A flexible asset can be used for optimization power consumption and generation across markets. In our project, an electrolyser is used to take positions in the day-ahead market and then offload them in the balancing market. Yet, positioning to capture the best price is an extremely difficult balancing act and is highly dependent on local conditions in a given zonal market. Therefore, we explore a RL-based algorithmic trading strategy to optimize the positioning in the day-ahead market and optimal bidding in the balancing market.

\section{Methodology}\label{Methods}

\subsection{Virtual Market Simulation Environment}
RL agents learn continuously from a simulation environment. Hence, the accuracy and efficiency of the environment directly impact the learning ability of the agent. We separate the environment into two levels, the day-ahead (DA) market level and the balancing market (BM) level. The DA agents observe the day-ahead features and decide the amount of power to be purchased in the DA market (the continuous action space is one-dimensional). Then, based on the observed BM features and the brought position and associated settlement price in the DA market, the BM agents output desired bid and ask prices for the BM bidding where at most one will be executed according to the regulatory state and settlement price range (the continuous action space is two-dimensional), e.g., the bid is executed in the surplus state, and the ask is executed in the shortage state if the submitted price is within the settlement range. A 3-day look-back window is added in both DA and BM market levels for agents to learn from history.

The DA market operates hourly, while the BM market operates in 15-minute intervals. To simulate a one-hour time slot, an episode is created, which consists of five steps: one DA step and four BM steps per quarter hour. In the first DA step, the initial purchase of electricity is deducted. Then, in the following BM steps, depending on the settlement range and regulatory state, the agent will sell all of the purchased electricity in the BM market, buy more BM positions and convert it into hydrogen which is in average sold to 75 Euro/MWh, or not operate in this interval and convert all remained DA position into hydrogen for profit. Then. The final hourly P\&L is the reward for this episode, shared by both agents. Additionally, due to a physical constraint from the electrolyzer, the agents must keep the operating power range of the electrolyzer between 20 MW and 200 MW, which limits the action space of the DA and BM agents. The key elements are outlined as follows:

{\bf Agents} In each hour, A DA agent decides the day-ahead position in advance for the entire hour slot, while a BM agent executes bidding in the balancing market for 4 15-minute intervals. The agents learn individually how to improve their day-ahead strategy and balance bidding by learning from repeated interactions with the virtual market environment. 

{\bf State Space} DA agent has 344 observation states and BM agent has 156 observation states corresponding to the actual and forecast features mentioned in Section~ \ref{Data} with associated lags, a detailed feature list is included in the Appendix \ref{App: Feature List}, where the column title matches exactly the raw data source in the TenneT databse.

{\bf Action Space} The action space of DA agent is a one-dimensional continuous space with a physical constraint from the electrolizer between 20 MWh to 200 MWh, representing the amount of the day ahead position bought, $s^{DA} \in [20,200]$. The action space of BM agent is a two-dimensional continuous space, $[P_b,P_a]$. We limit the bidding price between -200 and 200 $Euro/MWh$, where we have the balancing bid price $P_{b} \in [-200,200]$ and the balancing ask price $P_{a} \in [-200,200]$.

{\bf Reward Function} The DA agent and the BM agent are connected through the reward function, expressed by the P\&L in each hour or 15-min slot respectively. The BM agent receives a reward, $r^{BM}_{t}$, at $t^{th}$ 15-minute interval. which is defined as
\begin{equation}\label{eq:bm_reward}
\begin{aligned}
    r^{BM}_{t} =  (s^{BM}+s^{DA})\times p^{H} - s^{DA}\times p^{DA} - s^{BM}\times p^{BM}  
\end{aligned}
\end{equation}
where $s^{BM}$ is the position bought/sold in the Balancing Market, $p^{BM}$ is the associated execution price (i.e. an executed bid buys position and results in $s^{BM}>0$, and instead, an executed ask sells position and results $s^{BM}<0$),$p^{DA}$ is the price paid to buy position in the DA market ,and $p^{H}$ is the hydrogen price at which the remaining position is settled by generating green energy, hydrogen. Therefore, the reward is the net difference between the selling price and buying price in both DA and BM if the position has been cleared in BM, otherwise, all the remaining positions will be converged to hydrogen and settled.

For the DA agent, the reward at $T^{th}$ hour is simply the summation of the separate rewards in 4 15-min intervals, which is expressed as:
\begin{equation}\label{eq:da_reward}
\begin{aligned}
    r^{DA}_{T} =  \sum_{t\in T} {r^{BM}_{t}}
\end{aligned}
\end{equation}

\subsection{Reward Engineering} \label{IL}

The training of RL agents is computationally demanding and difficult, and agents can easily get stuck in local minima. Hence, to further improve the learning of the agents, we propose the reward engineering setup to leverage domain knowledge to guide training.

The initial rewards for BM and DA agents are defined by equation \ref{eq:bm_reward} and \ref{eq:da_reward}, which is simply the quarterly and hourly P\&L generated from each episode. The power system is intrinsically a physical system, heuristic methods can be constructed and serve as a baseline / benchmark. We construct the reward as the difference between the quarterly P\&L generated by the RL agents and the quarterly P\&L generated by the heuristic methods. This way, we incorporate additional domain knowledge as a prior in the reward function to guide the RL agents to learn outperform the baseline. Therefore the updated reward function for BM agent, denoted as $r^{BM}_{I,t}$ is:
\begin{equation}\label{eq:bm_reward_i}
\begin{aligned}
    r^{BM}_{I,t} =  r^{BM}_{t}- (F_{1}+F_{2}+F_{3})/3
\end{aligned}
\end{equation}
where, $F_{1}$, $F_{2}$ and $F_{3}$ are the quarterly P\&L generated the P1, P2 and P3 defined in table \ref{tab:bp}.

\subsection{Portfolio Tranching} \label{PT}

The BM market operates in an order-book-like auction structure, where bids and asks are settled in a centralized way at the end of each quarter-hour interval. Bids and asks outside of the current spread will not be executed and are discarded. To enhance the success rate of bidding, tranching is leveraged to break a large order into smaller orders distributed across the bid ladder to take advantage of the paid-as-bid nature of auctions. 

By allocating the volume from our day-ahead market position across the different price levels of the bid ladder we can treat this as a portfolio of assets. The volume is allocated for different price levels, $l$, spaced over 20 EUR/MWh, starting at 75 EUR/MWh and ending at 275 EUR/MWh for ask and starting at -125 EUR/MWh and ending at 75 EUR/MWh for bid. The choice of 75 EUR/MWh derives from the price received for hydrogen. Therefore, the reward function for BM agent, $r^{BM}_{P,t}$ is expressed as:

\begin{equation}\label{eq:bm_reward_p}
\begin{aligned}
    r^{BM}_{P,t} =  (\sum_{l} s^{BM}_{l}+s^{DA}) \times p^{H} -  s^{DA}\times p^{DA} - \sum_{l}{s^{BM}_{l} \times p^{BM}_{l} }
\end{aligned}
\end{equation}

where $s^{BM}_{l}$ and $p^{BM}_{l}$ represent the volume and price at price level $l$. Hence, the integrated reward engineering and portfolio tranching BM reward function,$r^{BM}_{IP,t}$, is defined as:

\begin{equation}\label{eq:bm_reward_pi}
\begin{aligned}
    r^{BM}_{IP,t} = r^{BM}_{P,t} - (F_{4}+F_{5})/2
\end{aligned}
\end{equation}
where $F_{4}$ and $F_{5}$ are the quarterly P\&L generated the P4 and P5 defined in table \ref{tab:bp}.

\section{Experiment Setup}\label{Exp}
\subsection{Data}\label{Data}

Settlement in power markets operates on a interval basis, where equity-spot-like continuous time series are absent. Therefore, technical analysis is hard to perform, and fundamental features are crucial to train and optimising RL agents. For the day-ahead market, we consider cross-border flows, TenneT (a transmission system operator in the Netherlands and in large part of Germany) forecasted Net Transfer Capacities (NTCs), TenneT forecasted solar and wind generation, TenneT forecasted load, as well as 24 h lagged day-ahead electricity prices, actual generation from biomass, gas, nuclear, solar, waste, wind (offshore/onshore), actual total load, actual NTCs and residual load. For the balancing market, since time slots are correlated on an hourly basis, only one-hour lags are considered for the lagged features mentioned above while balancing market bid and ask prices and the regulatory state are considered with 24-hour lags. In addition, balancing market prices and bidding volumes are provided to balancing market agents. All the raw features mentioned above are from the dataset which is open and publicly accessible from the TenneT database.

Besides raw features which are listed in Appendix \ref{App: Feature List}, we construct the probability of a shortage/surplus regulatory state as an additional feature based on logistic regression prediction from the 24-hour lagged actual load, 24-hour lagged generation by type, forecasted generation by wind/solar, forecasted load day-ahead, forecasted NTCs, 24-hour lagged actual NTCs, day-ahead price actuals, residual demand as well as a weekend Boolean.

\begin{table}[h]

\begin{center}\begin{tabular}{|c|c|c|}\hline 
{\bf Train} & {\bf Test} & {\bf Accuracy} \\\hline 
2015 & 2016 & 62.5\% \\\hline 
2015 & 2016 & 63.5\% \\\hline 
2015 & 2020 & 61.4\% \\\hline 
2017 & 2018 & 63.1\% \\\hline 
2017 & 2020 & 58.3\% \\\hline 
2018 & 2020 & 60.4\% \\\hline 
2020 & 2018 & 65.1\% \\\hline 
 
\end{tabular} 

\caption[State prediction with logistic regression]{Accuracy of walk forward logistic regression of prediction of shortage/surplus.}
	\label{tab:statePred1}
\end{center}

\end{table}

The performance of the prediction is presented in Table~\ref{tab:statePred1}. Curiously, the accuracy remains roughly constant despite the gap between the training and testing set. 

\subsection{dual-agent DDPG Implementation} 

We build two Deep Deterministic Policy Gradient (DDPG) \citep{Lillicrap2016ContinuousCW} agents for the dual-agent reinforcement learning framework. The bi-level problem is solved independently by optimizing the reward function, strategy P\&L in the hourly slot and quarterly intervals for the DA and BM agent respectively. However, as the hourly reward is the summation of the four quarterly rewards, the learning of the two agents are, thus associated with one objective function. For simplicity, for both agents, the actor and critic network employ a 2-layer Multi-layer Perceptron (MLP) of 64 and 32 hidden neurons with Batch Normalization and ReLU as the activation function. The detailed optimal hyper-parameter values of each agent during our experiments are listed below in Table \ref{tab:hp} for reference, where $\alpha_{critic}$ and $\alpha_{actor}$ is the learning rate for the critic and actor network.

\begin{table}[h!]

\begin{center}\begin{tabular}{|c|c|c|c|c|}
\hline 
{\bf Agent} & {\bf $\alpha_{critic}$}& {\bf $\alpha_{actor}$} & {\bf Optimiser} & {\bf Batch Size} \\\hline 
BM Agent & $ 0.0025$ & $ 0.00025$ & Adam & 64 \\\hline 
DA Agent & $ 0.0025$ & $ 0.00025$ & Adam & 64 \\\hline 
 
\end{tabular} 

\vspace{0.5pt}

\caption[HyperParameter]{Optimal Hyper-parameter values.
}
	\label{tab:hp}
\end{center}
\end{table}

\subsection{Walk Forward Optimization}
\begin{table*}[h!]

\begin{center}\begin{tabular}{|c|l|}\hline 
{\bf Benchmark} & {\bf Policy}  \\\hline 
\multirow {2}{*} {P1} & {DA position with 150 MWh,}\\
 {} & {and a fixed BM strategy of Bid at -100 Euro/MWh and Ask at 100 Euro/MWh} \\\hline 
\multirow {2}{*} {P2} & {Buy max position (200 MWh) and convert everything to hydrogen (no BM)} \\
{}&{}\\\hline 
\multirow {2}{*} {P3} & {Buy max position (200 MWh) if hydrogen price exceeds the DA price,} \\
{} & {else buy min position (20 MWh), and convert all positions to hydrogen (no BM)} \\\hline 
\multirow {2}{*} {P4} & {DA position is predicted by agent, and fixed BM strategy of equally weighted portfolio}  \\
{}&{}\\\hline
\multirow {2}{*} {P5} & {Buy max position (200 MWh) if the hydrogen price is greater than the DA price,} \\
{} & {else buy min position (20 MWh), and fixed BM strategy of an equally weighted portfolio (P2+P4)} \\\hline 
 
\end{tabular} 

\caption[Benchmark policies]{Table of benchmark policies.}\label{tab:bp}
\end{center}

\end{table*}
Walk forward optimization is a method used in finance to determine the optimal parameters for a trading strategy. The trading strategy is optimized with in-sample data for a time window in a data series. The remaining data is reserved for out of sample testing. A small portion of the reserved data following the in-sample data is tested and the results are recorded. The in-sample time window is shifted forward by the period covered by the out of sample test, and the process is repeated. Lastly, all of the recorded results are used to assess the trading strategy.
We run walk forward optimization to reduce overfitting in our optimizations. Overfitting in trading is the process of designing a trading system that adapts so closely to the noise in historical data that it becomes ineffective in the future. A walk forward optimization forces us to verify that we are adjusting our strategy parameters to signals in the past by constantly testing our optimized parameters from out-of-sample data. Inexperienced traders tend to spend a lot of time optimizing every parameter of the entire historical data, and then they proceed to trade on these “optimized” parameters. The objective function here is to maximize the reward function. 
The agent learns from in-sample data on what parameters to be selected and optimized to maximize the reward, based on the selected best parameters model performance is evaluated. In our experiments, we train the model with two years of sliding window data and test on consecutive years.

\subsection{Benchmarks}\label{bench}

In our experiments, we depict five benchmarks to compare the performance of our RL agents. These benchmarks are constructed based on market understanding. P1, P2 and P3 are simple benchmarks which buy a fixed DA position, and place bidding in the BM on a single price ladder at both Bid and Ask. P4 and P5, however, are portfolio benchmarks, while buying similarly a fixed DA position, they place bidding across a range of price levels. The detailed benchmark policies are summarized in Table \ref{tab:bp}.

\section{Results}\label{results}

\subsection{Decision Analytics}
\begin{figure*}[h]
	\centering
	\includegraphics[scale=0.5]{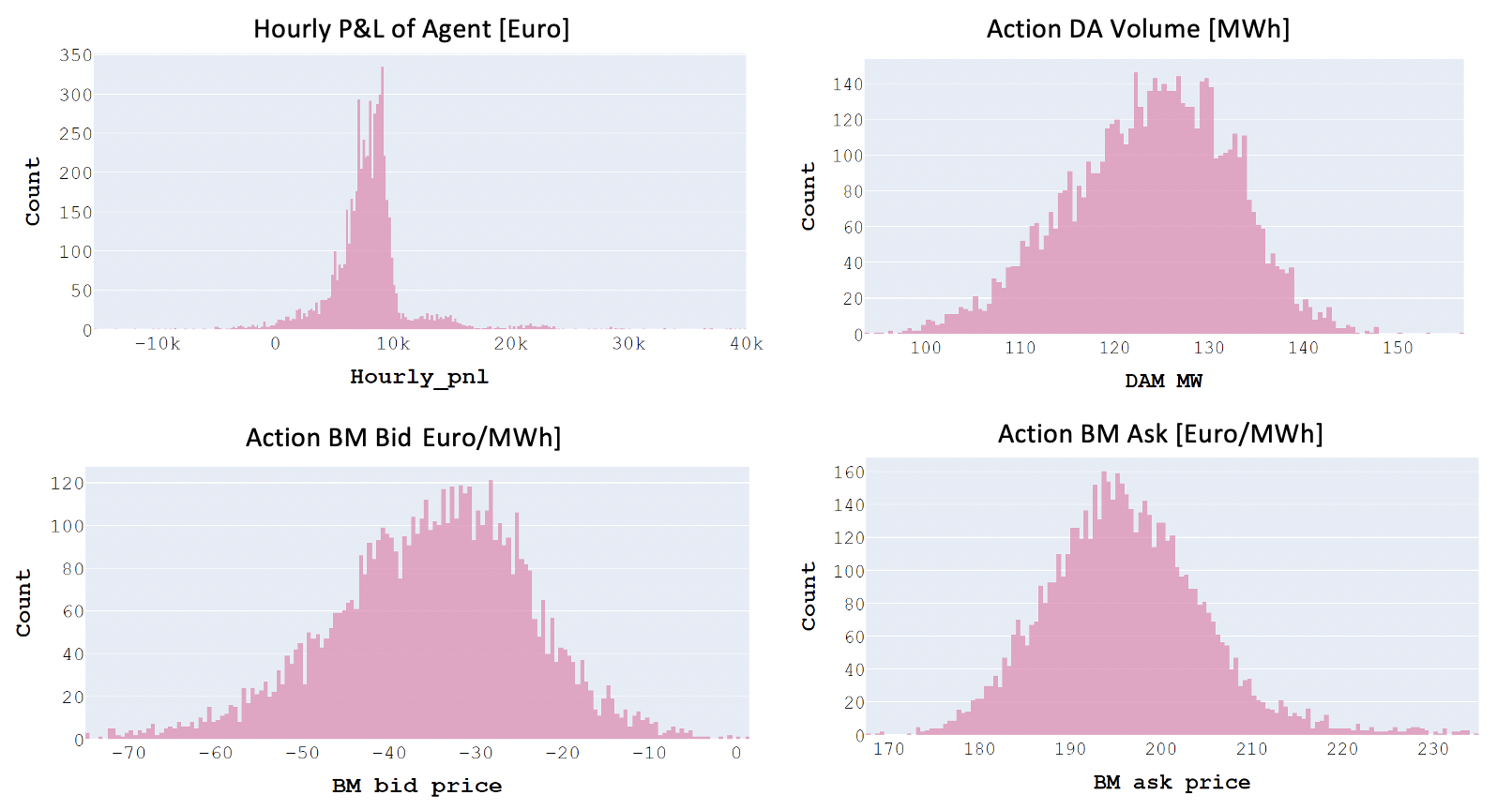}
	\vspace{0.5pt}
	\caption[Trade Characteristic Analysis]{Characteristic of trades and agent decisions in Walk Forward V1 (trained with 2015 \& 2016 and tested with 2017). The subfigures show a) the hourly P\&L, b) the distribution of agent DA actions, c) the distribution of agent bid BM actions, and d) the distribution of agent ask BM actions.}\label{fig:tradeAnalysis}
\end{figure*}

To analyze the series of decision made by the agent during trading and bidding, Figure~\ref{fig:tradeAnalysis} is presented outlining the ensemble statistics of hourly P\&L, DA volume as well as BM Bid/Ask decisions in histograms. Figure~\ref{fig:tradeAnalysis}.a shows the distribution of hourly cumulative P\&L of the agent. This negatively skewed distribution centres around 9000 with a long right tail, suggesting an average positive P\&L with occasional big gains on the right fat tail, which is often seen in other successful arbitrage trading systems \citep{Albuquerque2010SkewnessIS}.
Figure \ref{fig:tradeAnalysis}.b explains the distribution of agent DA actions. The maximum and minimum DA volume are constrained from 20 to 200 by the electrolyzer. This distribution roughly lies in the middle of the range, which shows extreme DA position are rarely taken by the agent unlike the defined benchmarks.

Figure~\ref{fig:tradeAnalysis}.c shows the distribution of agent bid market actions. Figure~\ref{fig:tradeAnalysis}.d shows the distribution of agent ask market actions. The two distributions show very profitable strategies taken by the agent in both bid and ask, compared to the hydrogen price at 75 EUR/MWh.

\subsection{Reward Engineering Result} \label{ilr}
\begin{figure}[h]
	\centering
	\includegraphics[scale=0.3]{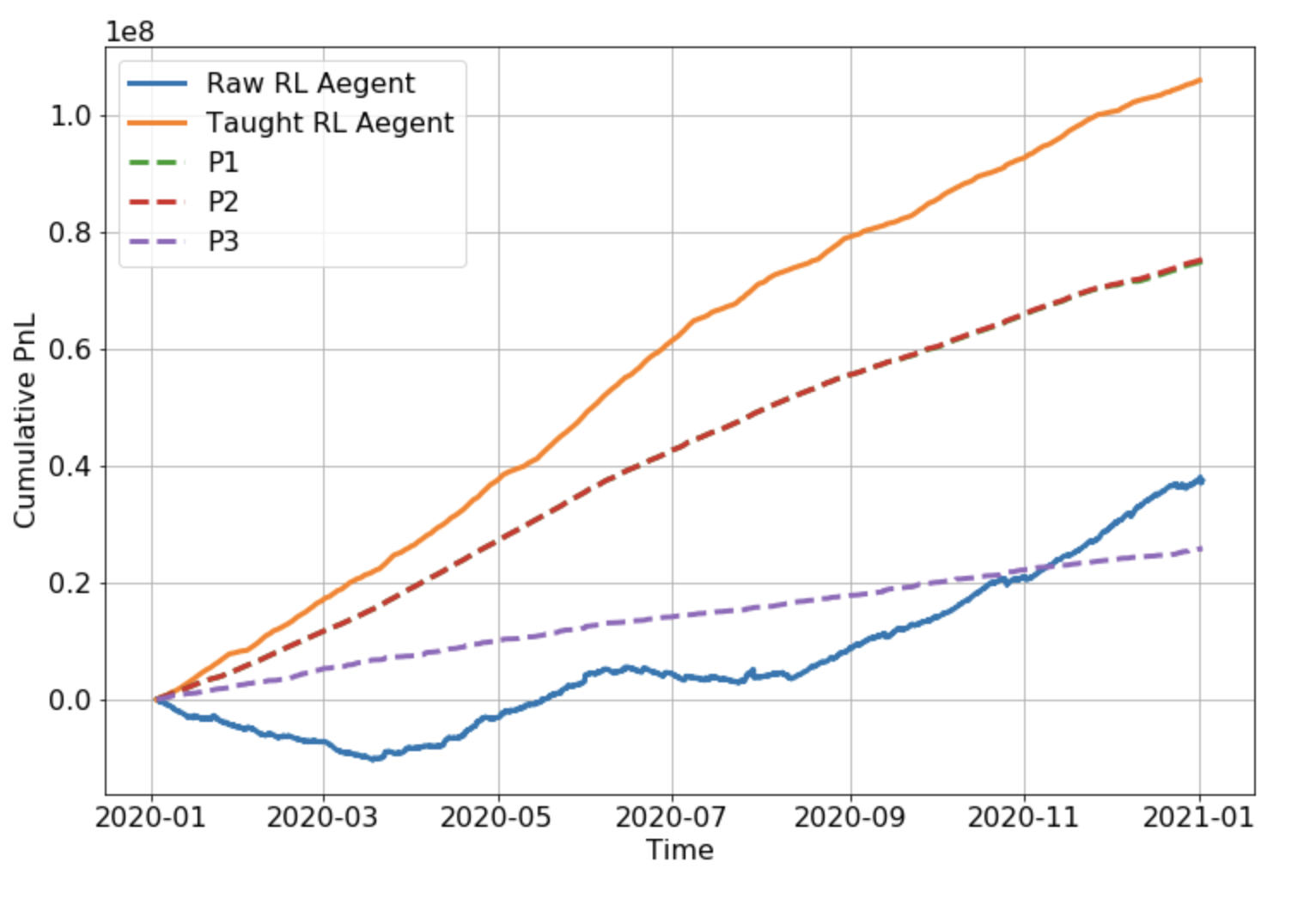}
	\vspace{0.5pt}
	\caption[In-house results with updated reward]{Comparison of agent P\&L against benchmarks, trained with 2018 \& 2019 and tested on 2020. P1, P2 and P3 are the benchmarks (P1 and P2 largely overlap). The Raw RL agent is the agent trained with standard reward function, and the Taught RL agent has a reconstructed reward as the difference between the standard reward and the performance of benchmarks.
		
	}\label{fig:inhouse1}
\end{figure}

Illustrated in Figure~\ref{fig:inhouse1} is the testing performance of our trained agent, marked as Raw RL Agent (blue), with respect to simple benchmarks (P1, P2 and P3). The agent is trained on 2018-2019 and tested on 2020. It is noticed that despite being profitable, the general performance of the Raw RL Agent is inferior to the three simple benchmarks constructed by our market understanding. The main reason for this inferiority is most likely resulted by the complexity of the market, and associated complicated bi-level problems simulated by our dual-agent setup. Therefore, to guide our agent to faster and better convergence in training, we leverage the knowledge in the three benchmarks as a prior for reward engineering. Detailed implementation is outlined in Section~\ref{IL}, and the testing performance is shown by the yellow line in Figure~\ref{fig:inhouse1}, denoted as Taught RL Agent.

The cumulative P\&L of the Raw RL Agent in 2020 is $38$M Euro, compared to P1 of $75$M Euro, P2 of $75$M Euro (P1 and P2 largely overlap) and P3 of $26$M Euro. It achieves about only a half of the two most performing benchmarks out of the three. However, after leveraging reward engineering techniques, the Taught RL Agent achieves $106$M Euro cumulative P\&L, which is about $40\%$ and $180\%$ improvement from the most performing benchmarks and the Raw RL Agent. Therefore, incorporating the knowledge from the benchmarks as a prior to guide learning in this scenario results in a significant uplifting in the performance.

\subsection{Portfolio Tranching Results} \label{Trahcnhing}

The Dutch balancing market is an order-book structural auction market. Hence, instead of placing entire volume of bidding at only one price level, we can also train the RL agent to split the volume into different levels of the book, termed portfolio tranching.

\begin{figure}[h]
	\centering
	\includegraphics[scale=0.3]{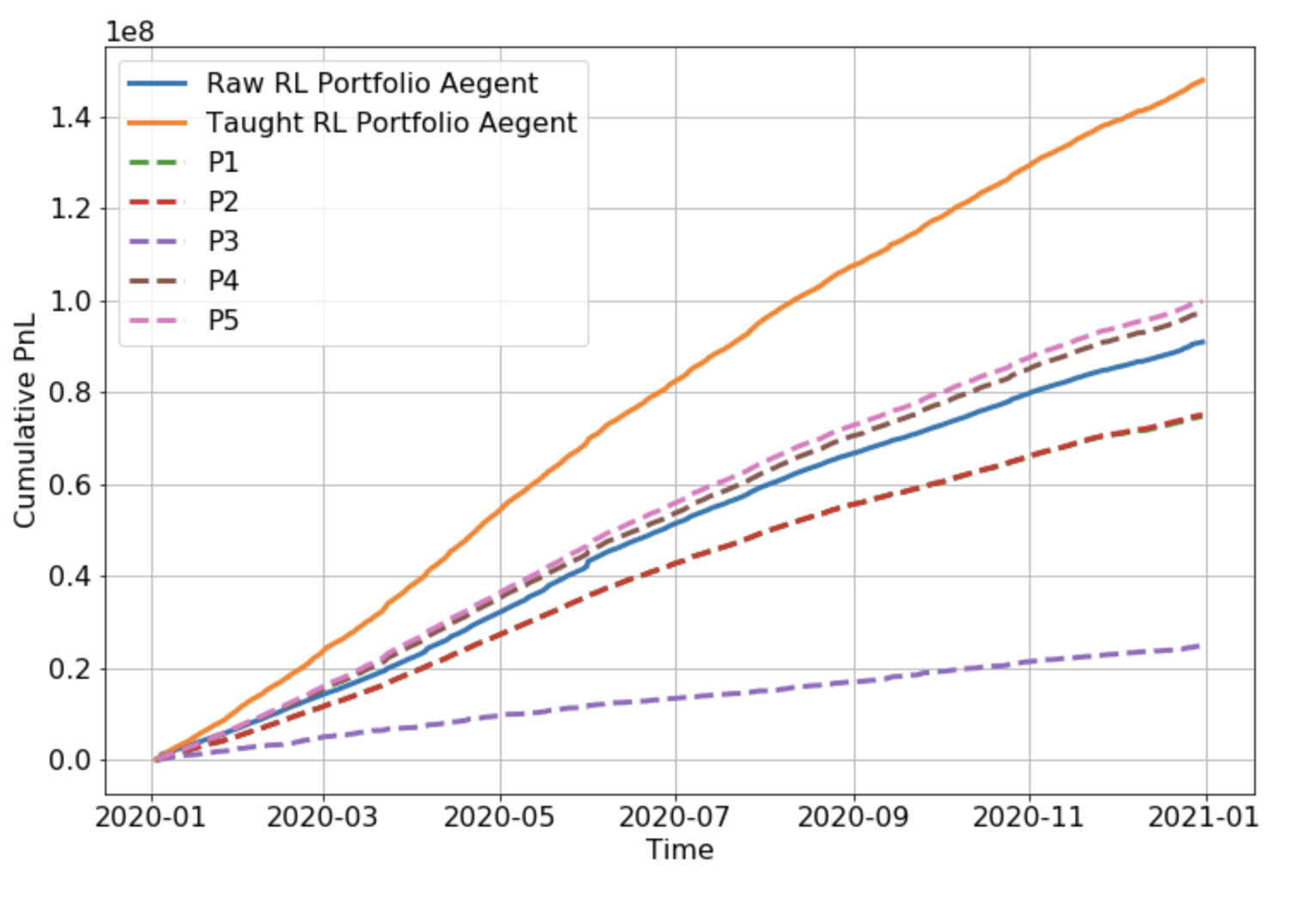}
	\vspace{0.5pt}
	\caption[Tranching Results]{Comparison of portfolio agent P\&L against benchmarks, trained with 2018 \& 2019 and tested on 2020. P4 and P5 are the portfolio benchmarks. The Raw RL Portfolio agent is the agent trained with standard reward function and portfolio tranching, and the Taught RL Portfolio Agent employs portfolio tranching with a reconstructed reward as the difference between the standard reward and the performance of P4 and P5.
		
	}\label{fig:inhouse2}
\end{figure}
In Figure~\ref{fig:inhouse2}, we demonstrate the performance of the Raw RL Portfolio Agents and the benchmarks. P1, P2 (P1 and P2 largely overlap) and P3 are the same simple benchmark reported in Figure~\ref{fig:inhouse1}, and P4 and P5 are portfolio benchmarks. The details of each benchmark are referenced in Section~\ref{bench}. It is notable that by employing portfolio tranching, even the Raw RL Portfolio Agent with standard reward function achieves $91$M Euro outperforming P1, P2 by about $20\%$, and naive Raw RL Agent in Section~\ref{ilr} by about $140\%$. This significant improvement illustrates the importance and efficacy of portfolio tranching.

Furthermore, in order to outperform the two portfolio benchmarks, P4 with cumulative P\&L of $97$M Euro and P5 with cumulative P\&L of $100$M Euro, we also leverage reward engineering on the portfolio tranching agent, denoted as Taught RL Portfolio Agent, which achieves cumulative P\&L of $148$M Euro. The guided learning method remarkably improve the cumulative P\&L by about $60\%$ from the Raw RL Portfolio Agent. Moreover, addictively incorporating both reward engineering and portfolio tranching methods, the cumulative P\&L is uplifted from $38$M Euro to $148$M Euro for about three-fold, while outperforms the highest benchmark P5 by about $50\%$. The notable results from this section clearly demonstrate the performance uplifting of the two practical implementations introduced by this paper, which also illustrates the importance of incorporating domain specific knowledge in RL problem settings.

\section{Conclusion}
\label{sec:Conclusion}

Power arbitrage trading is a complicated yet very profitable niche field in systematic commodity trading. The non-standard market structures and low transparency of the physical power grid settlement system prevent extensive exploration from the general quantitative finance community. In this paper, we present one of the first dual-agent Deep RL implementations for power arbitrage between the day-ahead market and the auction-like balancing market. The result is significant in that the cumulative P\&L of the RL agents outperform all the heuristic benchmarks.

Furthermore, we have explored two novel practical implementations to improve the training of the aforementioned framework. By leveraging the idea of reward engineering and restructuring the reward as the difference between the original reward and heuristic results, we incorporate additional domain information as a prior to guide agents to learn more effectively. This method has improved the performance of the naive RL agent and the portfolio RL agent by about $180\%$ and $60\%$ respectively.

Additionally, considering the order-book auction structural properties in the Dutch balancing power market, we introduce portfolio tranching to split a large order into equally weighted portfolio across several price levels. Not only does portfolio tranching improve the success bidding rate, moreover the performance has uplifted by about $140\%$. Hence, the combined final framework achieves an around three-fold improvement in cumulative P\&L, and outperforms the highest benchmark policy by around $50\%$.

Notably, in this paper, we have used a common RL algorithm (DDPG) with standard network configuration. Nevertheless, the two novel implementations inspired from domain specific knowledge and a representable virtual learning environment constructed from a realistic problem setting have boosted the results significantly. Therefore, from a practical point of view, the stable and performing deployment of RL does not need complicated state-of-art models, but an insightful settings and understanding of the problem.

This project has been a successful example of deep reinforcement learning implementation in power trading space. Further work is focusing on practical deployment and transforming this framework into other power markets across the globe.

\newpage
\begin{acks}
The authors would like to acknowledge the contributions of and collaboration with Simon Oliver (Shell), Tashi Erdmann (Shell) and Hossein Khadivi Heris (Microsoft Bons.ai).

Moreover, the authors would like to acknowledge the funding of the Shell.ai Futures Programme which has enabled this research and the Computational Sciences and Digital Innovations (CSDI) Lab for providing the compute resources (Azure).
\end{acks}
\newpage

\bibliographystyle{ACM-Reference-Format}
\bibliography{sample-base}
\newpage
\appendix
\section{Appendix}\label{App: Feature List}

\begin{table*}[h!]

\begin{center}

\begin{tabular}{|l|l|}
\hline
{\bf DA No-lag} &          {\bf DA 24h-lag} \\ \hline
                              BZN|BE $>$ BZN|NL  &                  Biomass  - Actual Aggregated  \\
                              BZN|NL $>$ BZN|BE  &               Fossil Gas  - Actual Aggregated  \\
                        BZN|DE-AT-LU $>$ BZN|NL  &         Fossil Hard coal  - Actual Aggregated  \\
                        BZN|NL $>$ BZN|DE-AT-LU  & Hydro Run-of-river and poundage   \\
                             BZN|DK1 $>$ BZN|NL  &                  Nuclear  - Actual Aggregated  \\
                             BZN|NL $>$ BZN|DK1  &                    Other  - Actual Aggregated  \\
                             BZN|NO2 $>$ BZN|NL  &                    Solar  - Actual Aggregated  \\
                             BZN|NL $>$ BZN|NO2  &                    Waste  - Actual Aggregated  \\
                              BZN|GB $>$ BZN|NL  &            Wind Offshore  - Actual Aggregated  \\
                              BZN|NL $>$ BZN|GB  &             Wind Onshore  - Actual Aggregated  \\
       Day-ahead Total Load Forecast  - BZN|NL &                    Actual Total Load  - BZN|NL \\
        Generation - Solar   Day Ahead/ BZN|NL &                          Day-ahead Price [EUR/MWh] \\
Generation - Wind Offshore   Day Ahead/ BZN|NL &                         BZN|BE $>$ BZN|NL Total  \\
 Generation - Wind Onshore   Day Ahead/ BZN|NL &                         BZN|NL $>$ BZN|BE Total  \\
                    BZN|BE $>$ BZN|NL Day Ahead  &                   BZN|DE-AT-LU $>$ BZN|NL Total  \\
                    BZN|NL $>$ BZN|BE Day Ahead  &                   BZN|NL $>$ BZN|DE-AT-LU Total  \\
              BZN|DE-AT-LU $>$ BZN|NL Day Ahead  &                        BZN|DK1 $>$ BZN|NL Total  \\
              BZN|NL $>$ BZN|DE-AT-LU Day Ahead  &                        BZN|NL $>$ BZN|DK1 Total  \\
                   BZN|DK1 $>$ BZN|NL Day Ahead  &                        BZN|NO2 $>$ BZN|NL Total  \\
                   BZN|NL $>$ BZN|DK1 Day Ahead  &                        BZN|NL $>$ BZN|NO2 Total  \\
                                                 &                         BZN|GB $>$ BZN|NL Total  \\
                                                 &                         BZN|NL $>$ BZN|GB Total  \\
\hline
\end{tabular}

\caption[DA FL]{DA Feature List with no lag an 24h lag}
	\label{tab:statePred1}
\end{center}

\end{table*}

\begin{table*}[h!]

\begin{center}

\begin{tabular}{|l|l|l|}
\hline
{\bf BM No-lag} &     {\bf BM 1h-lag} &      {\bf BM 24h-lag} \\
\hline
                              BZN|BE $>$ BZN|NL  &                  Biomass  - Actual Aggregated  & take\_from\_system\_EUR\_MWh \\
                              BZN|NL $>$ BZN|BE  &               Fossil Gas  - Actual Aggregated  & feed\_into\_system\_EUR\_MWh \\
                        BZN|DE-AT-LU $>$ BZN|NL  &         Fossil Hard coal  - Actual Aggregated  &         Regulation\_state \\
                        BZN|NL $>$ BZN|DE-AT-LU  & Hydro Run-of-river and poundage   &                        \\
                             BZN|DK1 $>$ BZN|NL  &                  Nuclear  - Actual Aggregated  &                        \\
                             BZN|NL $>$ BZN|DK1  &                    Other  - Actual Aggregated  &                        \\
                             BZN|NO2 $>$ BZN|NL  &                    Solar  - Actual Aggregated  &                        \\
                             BZN|NL $>$ BZN|NO2  &                    Waste  - Actual Aggregated  &                        \\
                              BZN|GB $>$ BZN|NL  &            Wind Offshore  - Actual Aggregated  &                        \\
                              BZN|NL $>$ BZN|GB  &             Wind Onshore  - Actual Aggregated  &                        \\
       Day-ahead Total Load Forecast  - BZN|NL &                    Actual Total Load  - BZN|NL &                        \\
        Generation - Solar   Day Ahead/ BZN|NL &                         BZN|BE $>$ BZN|NL Total  &                        \\
Generation - Wind Offshore   Day Ahead/ BZN|NL &                         BZN|NL $>$ BZN|BE Total  &                        \\
 Generation - Wind Onshore   Day Ahead/ BZN|NL &                   BZN|DE-AT-LU $>$ BZN|NL Total  &                        \\
                         Day-ahead Price [EUR/MWh] &                   BZN|NL $>$ BZN|DE-AT-LU Total  &                        \\
                    BZN|BE $>$ BZN|NL Day Ahead  &                        BZN|DK1 $>$ BZN|NL Total  &                        \\
                    BZN|NL $>$ BZN|BE Day Ahead  &                        BZN|NL $>$ BZN|DK1 Total  &                        \\
              BZN|DE-AT-LU $>$ BZN|NL Day Ahead  &                        BZN|NO2 $>$ BZN|NL Total  &                        \\
              BZN|NL $>$ BZN|DE-AT-LU Day Ahead  &                        BZN|NL $>$ BZN|NO2 Total  &                        \\
                   BZN|DK1 $>$ BZN|NL Day Ahead  &                         BZN|GB $>$ BZN|NL Total  &                        \\
                   BZN|NL $>$ BZN|DK1 Day Ahead  &                         BZN|NL $>$ BZN|GB Total  &                        \\
\hline
\end{tabular}

\caption[BM FL]{BM Feature List with no lag an 24h lag}
	\label{tab:statePred1}
\end{center}

\end{table*}

\end{document}